\documentclass{ws-p8-50x6-00}
\usepackage{amssymb}
\begin{document}

\title{Heavy quark production in $\gamma\gamma$ collisions}

\author{Ji\v{r}\'{\i} Ch\'{y}la}

\address{Center for Particle Physics, Institute of Physics, Academy
of Science of the Czech Republic, Na Slovance 2, 18221 Prague 8, Czech
Republic\\
E-mail: chyla@fzu.cz}

\maketitle
\abstracts{Current theoretical framework for the calculation
of heavy quark production in $\gamma\gamma$ collisions is
reviewed. The importance of including direct photon contributions
up to the order $\alpha^2\alpha_s^2$ and of the proper choice of
renormalization and factorization scales in the evaluation of
$\sigma(\gamma\gamma\rightarrow Q\overline{Q})$ is emphasized.}

\section{Introduction}
Heavy quark production in $\gamma\gamma$ collisions has recently
received increased theoretical attention \cite{zerwas,kramer,laenen}
motivated in part by new experimental data on $c\overline{c}$ and
$b\overline{b}$ production from LEP2 (see \cite{qqbar} for
references). Although the data on $b\overline{b}$ production
have sizable errors and the theoretical predictions suffer from
uncertainties, the excess by a factor of about three of the data over
theoretical predictions represents serious problem for perturbative
QCD. Interestingly, similar excess has been observed also in
$\gamma^*$p and $\overline{\mathrm{p}}$p collisions.

In such a situation it appears timely to reanalyze the theoretical
framework used for analyses of heavy quark production in $\gamma\gamma$
collisions \cite{qqbar}, paying particular attention to the
renormalization and factorization scale dependence, as these represent
the main source of the theoretical uncertainty.

\section{Basic facts}
The factorization scale dependence of
parton distribution functions (PDF) of the photon is determined by the
system of coupled inhomogeneous evolution equations
for quark singlet and nonsinglet and gluon distribution functions
\begin{eqnarray}
\frac{{\mathrm d}\Sigma(x,M)}{{\mathrm d}\ln M^2}& =&
\delta_{\Sigma}k_q+P_{qq}\otimes \Sigma+ P_{qG}\otimes G,
\label{Sigmaevolution}
\\ \frac{{\mathrm d}G(x,M)}{{\mathrm d}\ln M^2} & =& k_G+
P_{Gq}\otimes \Sigma+ P_{GG}\otimes G, \label{Gevolution} \\
\frac{{\mathrm d}q_{\mathrm {NS}}(x,M)}{{\mathrm d}\ln M^2}& =&
\delta_{\mathrm {NS}} k_q+P_{\mathrm {NS}}\otimes q_{\mathrm{NS}},
\label{NSevolution}
\end{eqnarray}
where, for $n_f$ massless quark flavors,
$\Sigma(x,M)\equiv \sum_{i=1}^{n_f}(q_i(x,M)+\overline{q}_i(x,M) )$,
$\delta_{\Sigma}=6n_f\langle e^2\rangle$ and $q_{\mathrm{NS}}(x,M)$
and $\delta_{\mathrm{NS}}$ can be defined in two different ways.
The splitting functions admit perturbative expansion
\begin{eqnarray}
k_q(x,M) & = & \frac{\alpha}{2\pi}\left[k^{(0)}_q(x)+
\frac{\alpha_s(M)}{2\pi}k_q^{(1)}(x)+
\left(\frac{\alpha_s(M)}{2\pi}\right)^2k^{(2)}_q(x)+\cdots\right],
\label{splitquark} \\ k_G(x,M) & = &
\frac{\alpha}{2\pi}\left[~~~~~~~~~~~~
\frac{\alpha_s(M)}{2\pi}k_G^{(1)}(x)+
\left(\frac{\alpha_s(M)}{2\pi}\right)^2k^{(2)}_G(x)+\cdots\;\right],
\label{splitgluon} \\ P_{ij}(x,M) & = &
~~~~~~~~~~~~~~~~~~\frac{\alpha_s(M)}{2\pi}P^{(0)}_{ij}(x) +
\left(\frac{\alpha_s(M)}{2\pi}\right)^2 P_{ij}^{(1)}(x)+\cdots.
\label{splitpij}
\end{eqnarray}
where $k_q^{(0)}(x)=(x^2+(1-x)^2)$ and $P^{(0)}_{ij}(x)$ are
{\em unique}, whereas all higher order ones
$k^{(j)}_q,k^{(j)}_G,P^{(j)}_{kl},j\ge 1$ depend  on the choice of
the {\em factorization scheme}. The equations
(\ref{Sigmaevolution}-\ref{NSevolution}) can be recast into evolution
equations for $q_i(x,M),\overline{q}_i(x,M)$ and $G(x,M)$ with
inhomogeneous splitting functions $k_{q_i}^{(0)}=3e_i^2k_q^{(0)}$.
The couplant $\alpha_s$ depends on the {\em renormalization scale}
$\mu$ and satisfies the equation
\begin{equation}
\frac{{\mathrm d}\alpha_s(\mu)}{{\mathrm d}\ln \mu^2}\equiv
\beta(\alpha_s(\mu))=
-\frac{\beta_0}{4\pi}\alpha_s^2(\mu)-
\frac{\beta_1}{16\pi^2}
\alpha_s^3(\mu)+\cdots,
\label{RG}
\end{equation}
where for $n_f$ massless quarks
$\beta_0=11-2n_f/3$ and $\beta_1=102-38n_f/3$.
General solution of the evolution equations
(\ref{Sigmaevolution}-\ref{NSevolution})
can be written as the sum of a particular solution of the full
inhomogeneous equations and a general solution, called
{\em hadron-like} (HAD), of
the corresponding homogeneous ones. A subset of the former resulting
from the resummation of contributions of diagrams describing multiple
parton emissions off the primary QED vertex
$\gamma\rightarrow q\overline{q}$ and vanishing at $M=M_0$, are called
{\em point-like} (PL) solutions. Due to the arbitrariness in the choice
of $M_0$ the separation
\begin{equation}
D(x,M)= D^{\mathrm {PL}}(x,M,M_0)+D^{\mathrm{HAD}}(x,M,M_0).
\label{separation}
\end{equation}
is, however, ambiguous.
Because of a different nature of the UV renormalization of the
couplant $\alpha_s(\mu)$ and the mass factorization involved
in the definition of dressed PDF I will keep the renormalization and
factorization scales $\mu$ and $M$ as independent free parameters.

In \cite{zerwas,kramer,laenen} the ``next--to--leading order''
QCD approximation to $\sigma(\gamma\gamma\rightarrow Q\overline{Q})$
is defined by taking into account the first two terms in the expansion
of direct, as well as single and double resolved photon contributions
\begin{eqnarray}
\sigma_{\mathrm{dir}} & = & \sigma_{\mathrm{dir}}^{(0)}+~~~~~
\sigma_{\mathrm{dir}}^{(1)}\alpha_s(\mu)+
\sigma_{\mathrm{dir}}^{(2)}(M,\mu)\alpha_s^2(\mu)+
\sigma_{\mathrm{dir}}^{(3)}(M,\mu)\alpha_s^3(\mu)+\cdots,~\label{dir}\\
\sigma_{\mathrm{sr}} & = & ~~~~~~~~
\sigma_{\mathrm{sr}}^{(1)}(M)\alpha_s(\mu)+
\sigma_{\mathrm{sr}}^{(2)}(M,\mu)\alpha_s^2(\mu)+
\sigma_{\mathrm{sr}}^{(3)}(M,\mu)\alpha_s^3(\mu)+\cdots, \label{sr}\\
\sigma_{\mathrm{dr}} & = & ~~~~~~~~~~~~~~~~~~~~~~~~~~~~~~~~
\sigma_{\mathrm{dr}}^{(2)}(M)\alpha_s^2(\mu)+
\sigma_{\mathrm{dr}}^{(3)}(M,\mu)\alpha_s^3(\mu)+\cdots. \label{dr}
\end{eqnarray}
Three light quarks were considered as intrinsic for $c\overline{c}$
and four in the case of $b\overline{b}$ production. Note that
$\sigma^{(0)}_{\mathrm{dir}}=\sigma_0c^{(0)}(s/m_Q^2)$,
where $\sigma_0\equiv 12\pi\alpha^2e_Q^4/s$ and $c^{(0)}(x)$ is
a function of $s/m_Q^2$ only, comes from pure QED.

The direct, single resolved and double resolved contributions as
considered in \cite{zerwas,kramer,laenen} start and end in
(\ref{dir}-\ref{dr}) at different powers of $\alpha_s$ . In the
conventional approach this is justified by claiming that PDF of the
photon behave as $\alpha/\alpha_s$ and, consequently, all three
expansions in fact  start and end at the same powers
$(\alpha_s)^0=1$ and $\alpha_s$, respectively. However, as argued in
\cite{factor}, the logarithm $\ln M^2$ characterizing the large $M$
behaviour
of PDF of the photon comes from integration over the transverse degree
of freedom of the purely QED vertex $\gamma\rightarrow q\overline{q}$
and cannot therefore be interpreted as $1/\alpha_s(M)$. If QCD is
switched off by sending, for fixed $M$ and $M_0$, $\Lambda\rightarrow 0$,
the quark and gluon distribution functions of the photon approach the
QED expressions
\begin{equation}
q_i(x,M)\rightarrow
q_i^{\mathrm{QED}}\equiv\frac{\alpha}{2\pi}3e_i^2k_q^{(0)}(x)
\ln\frac{M^2}{M_0^2},~
G(x,M)\rightarrow G^{\mathrm{QED}}=0.
\label{qQED}
\end{equation}
This is manifestly true for the point-like parts of quark and gluon
distribution functions, whereas the vanishing of the hadron-like parts
$q^{\mathrm{HAD}}$ and $G^{\mathrm{HAD}}$ as $\Lambda\rightarrow 0$
follows from the fact that vector mesons (or in general any hadronic
structure) can develop from the photon only as a consequence
of strong interactions between the $q\overline{q}$ pair coming
from the primary, purely QED, vertex $\gamma\rightarrow q\overline{q}$.
As QCD coupling vanishes, so must these hadronic parts.
Let me emphasize that there is no obstacle to performing this
limit, as by decreasing $\Lambda$ we get ever closer to the asymptotic
freedom point $\alpha_s=0$ and thus our perturbation expansions are
progressively better behaved. In the limit of vanishing colour coupling
we thus get the purely QED result $\sigma_{\mathrm{dir}}^{(0)}$. Had the
PDF of the photon really behaved as $\alpha/\alpha_s$, we would
get finite contributions from the lowest order single and double
resolved photon contributions (\ref{sr},\ref{dr}) even in the limit of
switching QCD off, which is clearly untenable.

All calculations \cite{zerwas,kramer,laenen} have been performed with
fixed quark masses, i.u. define $m_Q$ in the on-shell scheme. In this
convention $\sigma^{(1)}_{\mathrm{dir}}=\sigma_0c^{(1)}(s/m_Q^2)$, where
$c^{(1)}(x)$ is again a function of $s/m_Q^2$ only.

\section{Direct photon contribution}
For proper treatment of $\sigma_{\mathrm{dir}}$, the total cross
section for e$^+$e$^-$ annihilations into hadrons provides a suitable
guidance. For $n_f$ massless quarks we have
\begin{equation}
\sigma_{\mathrm{had}}(Q)=
             \sigma_{\mathrm{had}}^{(0)}+
\alpha_s(\mu)\sigma_{\mathrm{had}}^{(1)}(Q)+
\alpha_s^2(\mu)\sigma_{\mathrm{had}}^{(2)}(Q/\mu)+\cdots=
\sigma_{\mathrm{had}}^{(0)}(1+r(Q)),
\label{Rlarge}
\end{equation}
where $\sigma_{\mathrm{had}}^{(0)}\equiv
(4\pi\alpha^2/Q^2)\sum_fe_f^2$ comes,
similarly as $\sigma_{\mathrm{dir}}^{(0)}$ in (\ref{dir}), from pure
QED, whereas genuine QCD effects are contained in the quantity
\begin{equation}
r(Q)=\frac{\alpha_s(\mu)}{\pi}\left[1+\frac{\alpha_s(\mu)}
{\pi}r_1(Q/\mu)+\cdots\right].
\label{rsmall}
\end{equation}
In the case of the quantity (\ref{Rlarge}) nobody calls the lowest
order term $\sigma_{\mathrm{had}}^{(0)}(Q)$ the ``LO'' and the next
one, i.e. $\sigma_{\mathrm{had}}^{(0)}(Q)\alpha_s(\mu)/\pi$, the
``NLO'' QCD approximations, but these terms are reserved for genuine
QCD effects described by $r(Q)$! To work in a well-defined
renormalization scheme (RS) of $\alpha_s$ requires including in
(\ref{rsmall}) at least the first two (nonzero) powers of
$\alpha_s(\mu)$ because the dependence of the coefficients $r_k(\mu/Q)$
on the RS starts with $r_1$. The explicit $\mu$-dependence of
$r_1(Q/\mu)$ cancels to the order $\alpha_s^2$ the
implicit renormalization scale dependence of the leading order term
$\alpha_s(\mu)/\pi$ in (\ref{rsmall}) and thus guarantees that the
derivative of the sum of first two terms in (\ref{rsmall}) with
respect to $\ln \mu$ behaves as $\alpha_s^3$. Because $\alpha_s(\mu)$
is a monotonous function of $\mu$ spanning the whole interval
$(0,\infty)$, the inclusion of first two consecutive nonzero powers of
$\alpha_s$ is also a prerequisite for the applicability of any of the
scale fixing methods (see\cite{andrej} for detailed analysis of
renormalization scale dependence of (\ref{rsmall})). For purely
perturbative quantities the association of the term ``NLO QCD
approximation'' with a well-defined renormalization scheme is a
generally accepted convention, worth retaining for physical
quantities in any hard scattering process, like the direct photon
contribution $\sigma_{\mathrm{dir}}$ in (\ref{dir}).
\begin{figure}[t]
\epsfxsize=28pc 
\epsfbox{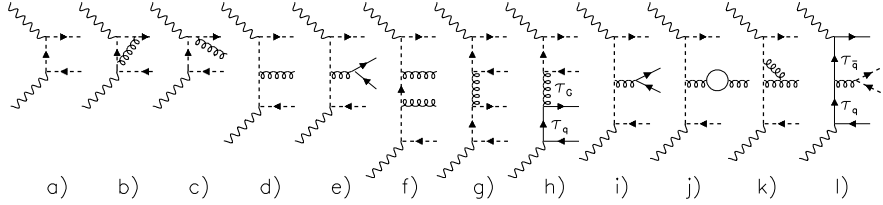} 
\caption{Examples of diagrams describing the direct photon contribution
to $\sigma(\gamma\gamma\rightarrow Q\overline{Q})$ up to the order
$\alpha^2\alpha_s^2$. The solid (dashed) lines denote light (heavy
quarks).}
\label{figdir}
\end{figure}

Unfortunately, in \cite{zerwas,kramer,laenen} the QED contribution
$\sigma_{\mathrm{dir}}^{(0)}$ is considered as the LO and the sum of
the first two terms in (\ref{dir}), which for fixed $m_Q$ has
exactly the same form as the first two term in (\ref{Rlarge}), as
the NLO approximation. Consequently, this approximation cannot be
associated to a well-defined renormalization scheme of $\alpha_s$
and therefore does not deserve the label ``NLO'' even if the NLO
expression for $\alpha_s(\mu)$ is used therein. For QCD analysis of
$\sigma_{\mathrm{dir}}$ in a well-defined renormalization scheme of
$\alpha_s(\mu)$ the incorporation of the third term in (\ref{dir}),
proportional to $\alpha^2\alpha_s^2$, is indispensable.

At the order $\alpha^2\alpha_s^2$ diagrams with light quarks appear as
well and we can thus distinguish three classes of direct photon
contributions differing by the overall heavy quark charge factor $CF$:
\begin{description}
\item{\bf Class A:} $CF=e_Q^4$. Comes from diagrams, like those in
Fig. \ref{figdir}a-g, in which both
photons couple to heavy $Q\overline{Q}$ pairs. Despite the presence of
mass singularities in contributions of individual diagrams coming from
gluons and light quarks in the final state and from loops,
at each order of $\alpha_s$ the sum of all contributions of this class
to $\sigma_{\mathrm{dir}}$ is finite.
\item{\bf Class B:} $CF=e_Q^2$.
Comes from diagrams, like that in Fig. \ref{figdir}h, in which one of
the photons couples to a heavy $Q\overline{Q}$ and the other to a light
$q\overline{q}$ pair. For massless light quarks this diagram has initial
state mass singularity, which is removed by introducing the concept of the
(light) quark and gluon distribution functions of the photon.
\item{\bf Class C:} $CF=1$. Comes from diagrams in which both
photons couple to light $q\overline{q}$ pairs, as those in Fig.
\ref{figdir}l. In this case the analogous subtraction procedure relates
it to the single resolved contribution of the diagram in Fig.
\ref{srpl}f and double resolved contribution of the
diagram in Fig. \ref{srpl}h.
\end{description}
As the diagrams in Fig. \ref{figdir}e and \ref{figdir}l give the
same final state $q\overline{q}Q\overline{Q}$, we have to consider their
interference term as well, but it turns out that it does not contribute
to the total cross section
$\sigma(\gamma\gamma\rightarrow Q\overline{Q})$.
Because of different charge factors, the classes A, B and
C do not mix under renormalization of $\alpha_s$ and factorization of
mass singularities. At the order $\alpha^2\alpha_s^2$ all three classes
of contributions are needed for theoretical consistency. As argued
above, class A is needed if the calculation of $\sigma_{\mathrm{dir}}$
is to be performed in a well-defined RS.

\section{Resolved photon contribution}
\begin{figure}[t]
\epsfxsize=28pc 
\epsfbox{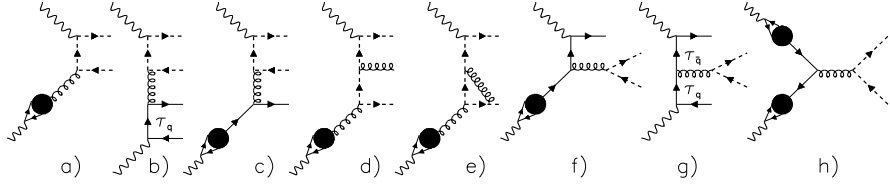} 
\caption{Examples of diagrams describing resolved and
related direct photon contributions.
\label{srpl}}
\end{figure}
The classes B and C of direct photon contributions are needed to render
the sum of direct and resolved photon contributions factorization
scale invariant. To see this, let us write the sum of first two terms in
(\ref{sr}-\ref{dr}) explicitly in terms of PDF and parton level cross
sections
\begin{eqnarray}
& &\sigma_{\mathrm{res}}^{(12)}(M,\mu)\equiv
2\alpha_s(\mu)\!\!\int\!\!\!{\mathrm{d}}xG(x,M)
\left[\sigma_{\gamma G}^{(1)}(x)+\alpha_s(\mu)\sigma_{\gamma G}^{(2)}
(x,M,\mu)\right]+
\label{p1}\nonumber\\
& &~~4\alpha_s^2(\mu)\!\!\int\!\!\!{\mathrm{d}}x\sum_iq_i(x,M)
\sigma_{\gamma q_i}^{(2)}(x,M)+\label{p44}\nonumber\\
& & ~~2\alpha_s^3(\mu)\!\!\int\!\!\!\!\int\!\!\!
{\mathrm{d}}x{\mathrm{d}}y\Sigma(x,M)G(y,M)
\sigma_{qG}^{(3)}(xy,M)+
\label{y1}\nonumber\\
& & ~~2\alpha_s^2(\mu)\!\!\int\!\!\!\!\int\!\!\!
{\mathrm{d}}x{\mathrm{d}}y\sum_i q_i(x,M)\overline{q}_i(y,M)
\left[\sigma_{q\overline{q}}^{(2)}(xy)+\alpha_s(\mu)
\sigma_{q\overline{q}}^{(3)}(xy,M,\mu)\right]+
\label{p4}\nonumber\\
& & ~~\alpha_s^2(\mu)\!\!\int\!\!\!\!\int\!\!\!
{\mathrm{d}}x{\mathrm{d}}yG(x,M)G(y,M)
\left[\sigma_{GG}^{(2)}(xy)+\alpha_s(\mu)\sigma_{GG}^{(3)}(xy,M,\mu)
\right],
\label{resolved}
\end{eqnarray}
where $\sum_iq_i$ runs over $n_f$ quark flavors and the factors of two
and four reflect the identity of beam particles and equality of
contributions from quarks and antiquarks.
Recalling the general form of the derivative
${\mathrm{d}}\sigma_{\mathrm{res}}/{\mathrm{d}}\ln M^2$
\begin{eqnarray}
\frac{{\mathrm{d}}\sigma_{\mathrm{res}}}{{\mathrm{d}}\ln M^2}
&=&\!\!\int\!\!{\mathrm{d}}x W_0(x,M)+\!\!
\int\!\!{\mathrm{d}}x\!\left[\sum_iq_i(x,M)W_{q_i}(x,M)+
G(x,M)W_G(x,M)\right]
\label{g2}\nonumber\\
& &\!\!\!\!\!\!\!\!\!\!\!\!\!\!\!\!
\!\!\!\!\!\!\!\!+\int\!\!\!\!\int\!\!{\mathrm{d}}x{\mathrm{d}}y
\!\!\left[G(x,M)G(y,M)W_{GG}(xy,M)\!+\!\sum_i
q_i(x,M)\overline{q}_i(y,M)W_{q\overline{q}}(xy,M)\right.\label{y9}
\nonumber\\
& & \left. +\Sigma(x,M)G(y,M)W_{qG}(xy,M) \right],
\label{derivace}
\end{eqnarray}
using (\ref{Sigmaevolution}-\ref{NSevolution}) and denoting
$\alpha_s\equiv\alpha_s(\mu)$,
$\dot{f}\equiv \mathrm{d}f/\mathrm{d}\ln M^2$ we find
\begin{eqnarray}
W_0&=&\frac{\alpha\alpha_s^2}{\pi}\left\{
\frac{1}{2\pi}k_G^{(1)}(x)\sigma_{\gamma G}^{(1)}(x)+
6k_q^{(0)}(x)\sum_i e_i^2\sigma_{\gamma q_i}^{(2)}(x,M)\right\}+
\cdots
\label{W0} \\
W_{q_i}&=& \frac{\alpha_s^2}{\pi}
\!\left\{4\pi\dot{\sigma}_{\gamma q}^{(2)}(x)+
\!\!\int\!\!{\mathrm{d}}z\!\left[P_{Gq}^{(0)}(z) \sigma_{\gamma
G}^{(1)}(xz)+3e_i^2\alpha k_q^{(0)}(z)
\sigma_{q\overline{q}}^{(2)}(xz)\right]\!\!\right\}\!\!+\cdots
\label{Wq}\\
W_G&=& \frac{\alpha_s^2}{\pi} \!\left\{2\pi\dot{\sigma}_{\gamma
G}^{(2)}(x)+ \!\!\int\!\!{\mathrm{d}}zP_{GG}^{(0)}(z)
\sigma_{\gamma G}^{(1)}(xz)\right\}+\cdots
\label{WG}\\
W_{GG}& = & \frac{\alpha_s^3}{\pi}
\left\{\pi\dot{\sigma}_{GG}^{(3)}(x)+\!\!\int\!\!
{\mathrm{d}}zP^{(0)}_{GG}(z)\sigma_{GG}^{(2)}(xz)\right\}+\cdots
\label{WGG}\\
W_{q\overline{q}}& = & \frac{\alpha_s^3}{\pi}
\left\{2\pi\dot{\sigma}_{q\overline{q}}^{(3)}(x)+2\!\!\int\!\!
{\mathrm{d}}zP^{(0)}_{qq}(z)\sigma_{q\overline{q}}^{(2)}(xz)\right\}
+\cdots\label{Wqq}\\
W_{qG}& = & \frac{\alpha_s^3}{\pi}
\left\{2\pi\dot{\sigma}_{qG}^{(3)}(x)+\!\!\int\!\!
{\mathrm{d}}z\!\left[P^{(0)}_{qG}(z)\sigma_{q\overline{q}}^{(2)}(xz)+
P^{(0)}_{Gq}(z)\sigma_{GG}^{(2)}(xz)\right]\!\right\}\!+\cdots
\label{WqG}
\end{eqnarray}
All integrals in (\ref{Wq}-\ref{WqG}) go formally from $0$ to $1$,
but threshold behaviour of cross sections $\sigma_{ij}(xz)$
restricts the region to $xz\ge 4m_Q^2/S$. The factorization scale
invariance of (\ref{resolved}) requires that the variation of
(\ref{resolved}) with respect to $\ln M^2$ is of higher order in
$\alpha_s$ than the approximation (\ref{resolved}) itself. There
is no question that direct photon contributions of classes B and C
are needed to make the sum direct and resolved photon
contributions factorization scale independent. The difference
between the conventional and my approaches to QCD analysis of
$\sigma(\gamma\gamma\rightarrow Q\overline{Q})$ concerns the
question when this happens, i.e what is the order of the
approximation (\ref{resolved}) and, consequently, which terms on
the r.h.s. of (\ref{derivace}) must vanish.

In the conventional approach both $q(M)$ and $G(M)$ are claimed to
be of order $\alpha/\alpha_s$, and the approximation (\ref{resolved})
thus of the order $\alpha^2\alpha_s$. This implies that only terms
up to this order must vanish in (\ref{derivace}). As the
expressions in round brackets of (\ref{Wq}) and (\ref{WG}) do, indeed,
vanish, the expression (\ref{resolved}) is claimed to be complete NLO
approximation to $\sigma_{\mathrm{res}}$. The fact that the expression
on the r.h.s. of (\ref{W0}) does not vanish is of no concern in this
approach as it is manifestly of the order $\alpha\alpha_s^2$ and thus
supposedly of
higher order than (\ref{resolved}) itself. This line of argumentation
is, however, untenable on physical grounds as it requires that not
only the quark but even the gluon distribution function $G(M)$ should
behave as $\alpha/\alpha_s$, despite the fact that gluons appear in the
photon only due to their radiation off the quarks and their contribution
to physical observables must therefore vanish if this radiation is
switched off!

If we discard the untenable claim that $q,G\propto
\alpha/\alpha_s$ and realize the obvious, namely that $q(M)\propto\alpha$,
we are lead to the conclusion that $W_0$ is of the same
order $\alpha^2\alpha_s^2$ as $q_iW_{q_i}$ and other integrands in
(\ref{derivace}) and must also vanish for theoretical consistency
of the approximation (\ref{resolved}). This, in turn, necessitates
the inclusion of class B direct photon contributions of the order
$\alpha\alpha_s^2$, like those in Fig. \ref{srpl}b,g, which
provide the $M$-dependent terms the derivative of which cancel
$W_0$ in the expression for
${\mathrm{d}}(\sigma_{\mathrm{dir}}(M)+
\sigma_{\mathrm{res}}(M))/{\mathrm{d}}\ln M^2$. Note that $W_{q_i}$
in (\ref{Wq}) receives contributions from the derivatives of both
single and double resolved photon diagrams, proportional to
$\sigma_{\gamma G}$ and $\sigma_{q\overline{q}}$, respectively.
This fact reflects the mixing of single and double resolved photon
contributions, which starts just at the order
$\alpha^2\alpha_s^2$. Note also, that for theoretical consistency
of the sum $\sigma_{\mathrm{dir}}+\sigma_{\mathrm{res}}$ up to
this order, only the lowest order double resolved photon
contribution needs to be included.

\section{Phenomenological implications}
Although the current calculations of
$\sigma(\gamma\gamma\rightarrow Q\overline{Q})$ do not represent
complete NLO QCD approximation, one may ask how important are the
missing terms, discussed above, numerically. To help answer this
question it would useful if the data could be expressed in terms
of energy dependence of $\sigma(\gamma\gamma\rightarrow
Q\overline{Q})$. The point is that for $b\overline{b}$ production
direct photon contribution dominates below about $\sqrt{s}\simeq 25$
GeV, whereas for
$200\gtrsim \sqrt{s}\gtrsim 50$ GeV single resolve photon
contribution makes up most of $\sigma(\gamma\gamma\rightarrow
b\overline{b})$. If the disagreement comes from the former region,
we would be in real trouble. As, however, the LEP data
integrate over the whole WW spectrum, to which the direct and
single resolved photon channels contribute roughly equally, there
is no way how to determine wherefrom comes the disagreement with
the data.

As argued above, the approximation used in
\cite{zerwas,kramer,laenen} for the evaluation of the direct
contribution is merely of the LO nature. On the other hand, as the
lowest order QCD correction $\sigma_{\mathrm{dir}}^{(1)}\alpha_s$
makes up a small fraction of the purely QED contribution
$\sigma_{\mathrm{dir}}^{(0)}$, the eventual incorporation of the
(so far unknown) direct photon term
$\sigma_{\mathrm{dir}}^{(2)}\alpha_s^2$, coming from class A direct
photon diagrams, will unlikely enhance
$\sigma_{\mathrm{dir}}$ by more than further few tens of percent.

The situation is different for the resolved photon contributions.
In \cite{next} I will discuss numerical aspects of the
factorization and renormalization scale dependence of the
approximation (\ref{resolved}) and show that proper choice of
these scales is essential for understanding the data on
$b\overline{b}$ production and may explain part of the observed
discrepancy between the data and theoretical calculations.

\section*{Acknowledgments}
The work has been supported by the Ministry of Education of the
Czech Republic under the project LN00A006.


\begin{thebibliography}{99}
\bibitem{zerwas} M. Drees, M. Kr\"{a}mer, J. Zunft, and P. Zerwas,
{\em Phys. Lett.} B {\bf 306}, 371
\bibitem{kramer}M. Kr\"{a}mer and E. Laenen,
\Journal{\PLB}{371}{303}{1996}.
\bibitem{laenen}S. Frixione, M. Kr\"{a}mer and E. Laenen,
\Journal{\NPB}{571}{169}{2000}.
\bibitem{qqbar} J. Ch\'{y}la, hep-ph/0010140
\bibitem{factor} J. Ch\'{y}la, JHEP04(2000)007.
\bibitem{andrej} J. Ch\'{y}la, A. Kataev and S. Larin,
\Journal{\PLB}{267}{269}{1991}
\bibitem{next} J. Ch\'{y}la, in preparation
\end{thebibliography}
\end{document}